\PassOptionsToPackage{unicode}{hyperref}
\PassOptionsToPackage{hyphens}{url}
\documentclass[
  12pt,
  a4paper,
]{article}
\usepackage{amsmath,amssymb}
\usepackage{lmodern}
\usepackage{iftex}
\ifPDFTeX
  \usepackage[T1]{fontenc}
  \usepackage[utf8]{inputenc}
  \usepackage{textcomp} 
\else 
  \usepackage{unicode-math}
  \defaultfontfeatures{Scale=MatchLowercase}
  \defaultfontfeatures[\rmfamily]{Ligatures=TeX,Scale=1}
\fi
\IfFileExists{upquote.sty}{\usepackage{upquote}}{}
\IfFileExists{microtype.sty}{
  \usepackage[]{microtype}
  \UseMicrotypeSet[protrusion]{basicmath} 
}{}
\makeatletter
\@ifundefined{KOMAClassName}{
  \IfFileExists{parskip.sty}{%
    \usepackage{parskip}
  }{
    \setlength{\parindent}{0pt}
    \setlength{\parskip}{6pt plus 2pt minus 1pt}}
}{
  \KOMAoptions{parskip=half}}
\makeatother
\usepackage{xcolor}
\usepackage[margin=1in]{geometry}
\usepackage{longtable,booktabs,array}
\usepackage{calc} 
\usepackage{etoolbox}
\makeatletter
\patchcmd\longtable{\par}{\if@noskipsec\mbox{}\fi\par}{}{}
\makeatother
\IfFileExists{footnotehyper.sty}{\usepackage{footnotehyper}}{\usepackage{footnote}}
\makesavenoteenv{longtable}
\usepackage{graphicx}
\makeatletter
\def\maxwidth{\ifdim\Gin@nat@width>\linewidth\linewidth\else\Gin@nat@width\fi}
\def\maxheight{\ifdim\Gin@nat@height>\textheight\textheight\else\Gin@nat@height\fi}
\makeatother
\setkeys{Gin}{width=\maxwidth,height=\maxheight,keepaspectratio}
\makeatletter
\def\fps@figure{htbp}
\makeatother
\newlength{\cslhangindent}
\newlength{\csllabelwidth}
\newlength{\cslentryspacingunit} 
\newenvironment{CSLReferences}[2] 
 {
  \setlength{\parindent}{0pt}
  \ifodd #1
  \let\oldpar\par
  \def\par{\hangindent=\cslhangindent\oldpar}
  \fi
  \setlength{\parskip}{#2\cslentryspacingunit}
 }%
 {}
\usepackage{calc}

\usepackage{mathptmx}
\usepackage{fancyhdr}

\usepackage{titling}
\usepackage{bm}
\pretitle{\begin{center}\LARGE }
\posttitle{\end{center}}
\ifLuaTeX
  \usepackage{selnolig}  
\fi
\IfFileExists{bookmark.sty}{\usepackage{bookmark}}{\usepackage{hyperref}}
\IfFileExists{xurl.sty}{\usepackage{xurl}}{} 
\hypersetup{
  pdftitle={Bayesian non-parametric specification of bathtub shaped hazard rate functions},
  pdfauthor={Richard Arnold\^{}1, Stefanka Chukova\^{}1 and Yu Hayakawa\^{}2},
  hidelinks,
  pdfcreator={LaTeX via pandoc}}

\title{Bayesian non-parametric specification of bathtub shaped hazard rate functions}
\author{Richard Arnold\(^1\), Stefanka Chukova\(^1\) and Yu Hayakawa\(^2\)}
\date{14 May 2023}

\begin{document}
\maketitle

\(^1\)School of Mathematics and Statistics, Victoria University of Wellington, New Zealand,\newline
\href{mailto:richard.arnold@vuw.ac.nz}{\nolinkurl{richard.arnold@vuw.ac.nz}}, \href{mailto:stefanka.chukova@vuw.ac.nz}{\nolinkurl{stefanka.chukova@vuw.ac.nz}}.

\(^2\)School of International Liberal Studies, Waseda University, Tokyo, Japan,\newline
\href{mailto:yu.hayakawa@waseda.jp}{\nolinkurl{yu.hayakawa@waseda.jp}}.

\hypertarget{abstract}{%
\subsubsection*{Abstract}\label{abstract}}
\addcontentsline{toc}{subsubsection}{Abstract}

Hazard rate functions of natural and manufactured systems often
show a bathtub shaped failure rate. A high early rate of failures
is followed by an extended period of useful working life where
failures are rare, and finally the failure rate increases as the system
reaches the end of its life. Parametric modelling of such hazard
rate functions can lead to unnecessarily restrictive assumptions on
the function shape, however the most common non-parametric estimator
(the Kaplan-Meier estimator) does not allow specification of the
requirement that it be bathtub shaped. In this paper we extend the
Lo and Weng (1989) approach and specify
four non-parametric bathtub hazard rate functions drawn from Gamma
Process Priors. We implement and demonstrate simulation for
these four models.

\textbf{Keywords:}
Reliability, Bayesian non-parametrics, Gamma Process, Bathtub hazard rate function.

\begin{table}[htbp]
\caption{Notation\label{tab:notation}}
\medskip
\begin{center}
\begin{tabular}{lp{0.7\linewidth}}
\hline
$\alpha$      & Shape parameter of the Gamma Process Prior\\
$a$           & Symmetry point of Lo-Weng Bathtub\\
$a_1, a_2$    & Parameters of hyperprior for $\alpha$\\
$\beta$       & Scale parameter of the Gamma Process Prior\\
$b_1, b_2$    & Parameters of hyperprior for $\beta$\\
$\text{DPP}(\alpha H_0)$ & Dirichlet Process Prior\\
$\phi$        & Parameter of exponential base probability measure $H_0$\\
$f_1, f_2$    & Parameters of hyperprior for $\phi$\\
$\text{Beta}(\cdot,\cdot)$ & Beta distribution\\
$\text{Exp}(\cdot)$ & Exponential distribution\\
$G(\cdot)$    & Random measure on $\Theta$\\
$G_K(\cdot)$  & Realisation of the truncated stick-breaking construction
                of $\text{GaPP}_K(\alpha H_0,\beta)$\\
$\text{Ga}(\cdot,\cdot)$ & Gamma distribution\\
$\text{GaPP}(\alpha H_0,\beta)$ & Gamma Process Prior\\
$\gamma$      & Total mass of measure $G(\Theta)$\\
$H_0(\cdot)$  & Probability measure on $\Theta$\\
$\text{IHPP}(\lambda(\cdot))$ & Inhomogeneous Poisson Process with 
                hazard rate function $\lambda(\cdot)$\\
$\kappa(t|u)$ & Mixing kernel\\
$K$           & Truncation limit for the stick-breaking construction\\
$\lambda(t|\cdot)$ & Hazard rate function\\
$\Lambda(t|\cdot)$ & Cumulative hazard rate function\\
$n$                & Total sample size\\
$n_0$              & Number of uncensored observations\\
$t$                & Failure time\\
$\tau$             & Censoring time\\
$\Theta$           & Set on which random measure $G(\cdot)$ is defined\\
$\bm{\theta}$      & Set of $K$ support points\\
$\theta_k$         & Location of the $k^{\rm th}$ support point\\
$\textbf{v}$       & Set of $K-1$ stick-breaking values\\
$v_k$              & $k^{\rm th}$ stick-breaking value\\
$w_k$              & Weight of the $k^{\rm th}$ support point\\
$\tilde{w}_k$      & Unscaled weight of the $k^{\rm th}$ support point\\
\hline
\end{tabular}
\end{center}
\end{table}

\hypertarget{sec:introduction}{%
\section{Introduction}\label{sec:introduction}}

Hazard rate functions of natural and manufactured systems often show a bathtub shaped
failure rate. A high early rate of failures is followed by an extended period of useful
working life where failures are rare, and finally the failure rate increases as the system
reaches the end of its life.

\begin{figure}

{\centering \includegraphics{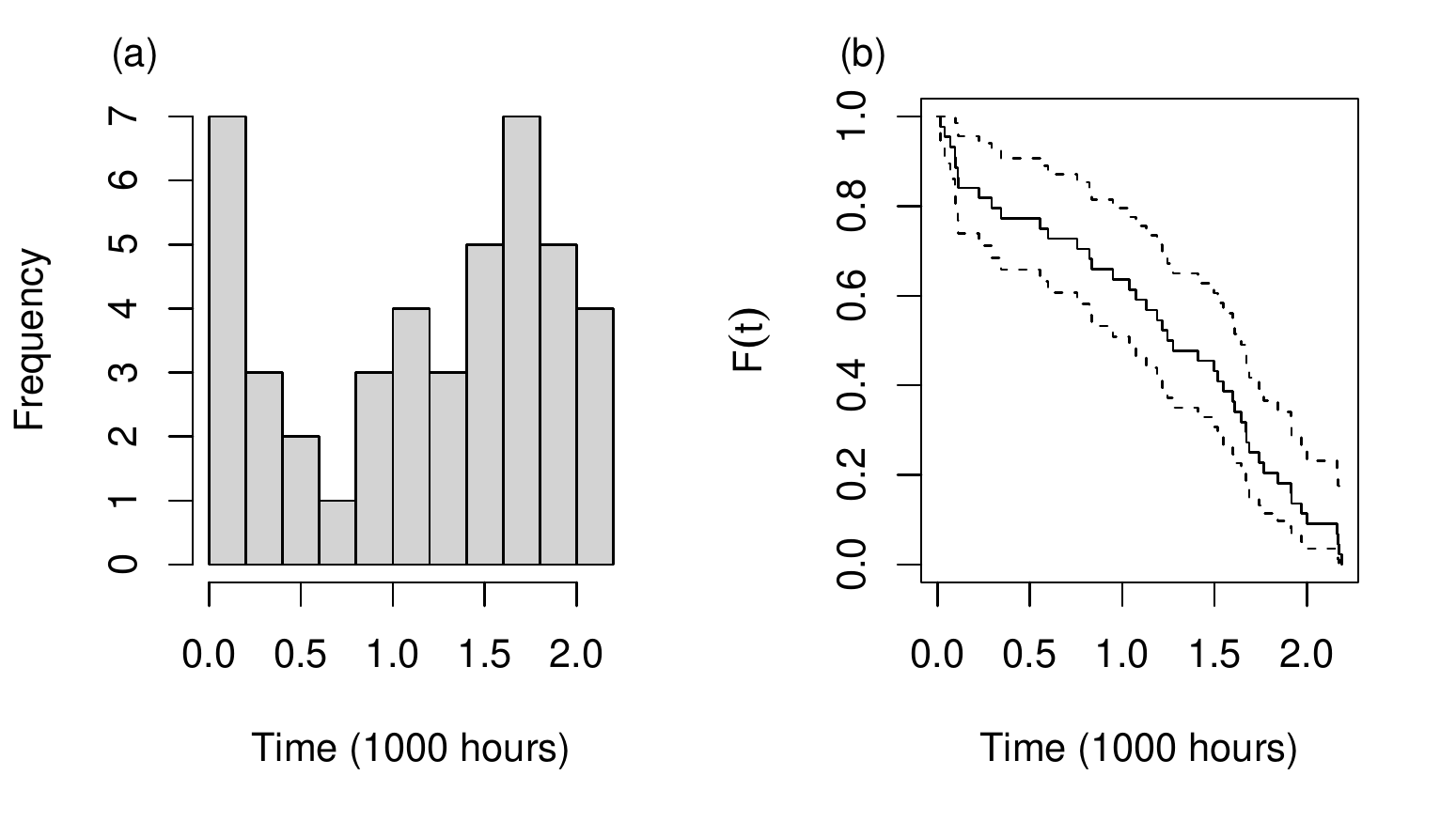} 

}

\caption{(a) Failure times of Load Hauling Machine A from Kumar, Klefsjö, and Granholm (1989). (b) Kaplan-Meier survival function of the machine failure times.}\label{fig:kumarLHD-A}
\end{figure}

An example of such data is shown in Figure \ref{fig:kumarLHD-A},
where a histogram of failure times of one Load-Haul-Dump machine from a Swedish
mine (Kumar, Klefsjö, and Granholm 1989). The associated empirical survival function
is plotted alongside. In this example the system is repairable, and
subject to multiple failures. These are more frequent earlier and later in the machine's
working life, associated with steep decline in the survival function in the early
and late periods, and a flatter decline between.

Data of this nature can also arise from observations of a set of independent
items, where the length of the useful working life of each item is recorded.
In this setting there is a population of items, some proportion of which
have manufacturing defects, and fail early. The remainder have low failure probability
until late times, at which point they begin failing in large numbers.

In either scenario the hazard rate starts high, decreases to some minimum, and
then climbs again. A `U-shaped' hazard rate is one where the low hazard rate
period is non-existent or very short, and a `bathtub shaped' hazard rate is one where
there is an extended period where the failure rate is low.

In both cases, the modelling of the associated hazard rate function may be carried out in
a number of ways. Inserting a polynomial dependence on time into simpler baseline failure distributions
is one approach, and another is to use a two component mixture of parametric distributions
(Glaser 1980). The three phases of the bathtub function can also be built from a piecewise
combination of parametric functions
(e.g. Peng, Liu, and Wang 2016). A further approach is to exponentiate a baseline failure distribution
(e.g. Abbasi et al. 2019; Iqbal et al. 2021). Although such approaches allow a wide range of behaviours,
parametric modelling can lead to unnecessarily restrictive assumptions on the hazard rate function shape.
Conversely, the most common non-parametric estimator of the hazard rate function
(the Kaplan-Meier estimator, Kaplan and Meier (1958)) is not sufficiently constrained, in that it
does not allow specification of the requirement that it be
bathtub shaped. In this paper we specify and investigate non-parametric bathtub hazard rate
functions based on the Gamma Process Priors, as described by Lo and Weng (1989). Although the Gamma
Process Prior does embed a parametric baseline failure distribution, it allows the posterior
distribution of the failure distribution to depart significantly from that baseline.

The Gamma Process has been applied to reliability in many settings (e.g Dykstra and Laud 1981),
and in particular it has been used as a conjugate prior for hazard rate functions.
Müller et al. (2015) and Phadia (2016) provide relevant introductions
to Bayesian non-parametrics. The theory of completely random measures by
Kingman (1967) led to the development of the Dirichlet Process Prior
(Ferguson 1974), and in particular its implementations in the Stick-Breaking
and the Chinese Restaurant Processes (Müller et al. 2015; Sethuraman 1994; Paisley 2010).\\
These advances, and the Stick-Breaking representations of the Beta (Paisley et al. 2010; Paisley, Blei, and Jordan 2012) and Gamma Processes (Roychowdhury and Kulis 2014, 2015)
have made Bayesian non-parametric approaches feasible for inference problems.

The work by Dykstra and Laud (1981) in Bayesian non-parametric approaches to reliability
defined a prior over hazard rates using the extended gamma process and were able
to derive posterior distributions for monotonic hazard rate function in analytic form.
Lo and Weng (1989) used a variety of kernel functions to weight a draw from a Gamma Process
and thus generate hazard rate functions of various forms. The weighted gamma process
was also used by Ho and Lo (2001) as a prior for hazard rate functions, and they implemented
a Markov Chain Monte Carlo method for deriving posterior properties of the hazard rate.

Hayakawa et al. (2001) used the Lo and Weng (1989) formulation to set up a Bayesian hypothesis test
for non-decreasing hazard rates. They also used a Monte Carlo method, based on the
weighted Chinese restaurant process (Lo, Brunner, and Chan 1998).

Arnold, Chukova, and Hayakawa (2020) demonstrated the Lo and Weng (1989) approach for hazard rate functions
that were increasing (IFR) using the Gamma Process Prior, and implemented a fully Bayesian non-parametric
approach to inference for such functions. In this paper we extend the results of that paper to
four specifications of bathtub failure rate hazard rate functions, all based on
Gamma Process Priors. We use a gamma-scaled Dirichlet
Process prior to implement the Gamma Process prior, and demonstrate the
methodology and the properties of the models using simulations.

Section \ref{sec:model} introduces the basic concepts of reliability modelling, the Gamma
Process Prior, and then lays out the definitions of six hazard rate functions.
Section \ref{sec:simulation} gives specific details of models conditional on
draws from the Gamma Process Prior, and protocols for simulation. We also
specify the priors required for each model. Section \ref{sec:demonstration} demonstrates
the properties of the models, showing specific draws from the Gamma Process Prior
and the properties of the hazard rate functions and the failure time distributions that
result. Brief concluding remarks close the paper in
Section \ref{sec:conclusions}, including a proposal for future work, including inference,
which we will carry out in a subsequent paper. We list our notation in Table \ref{tab:notation}.

\hypertarget{sec:model}{%
\section{Model}\label{sec:model}}

\hypertarget{sec:basicmodel}{%
\subsection{Basic Model Specification}\label{sec:basicmodel}}

We assume that a common hazard rate function \(\lambda(t)\) generates failures in
\(n\) identical items. The failure time distribution has
density \(f(t|\lambda(\cdot))\) and cumulative distribution function \(F(t|\lambda(\cdot))\).
The cumulative hazard function is
\begin{equation}
   \Lambda(t) = \int_0^t \lambda(u)\,{\rm d}u\ .
   \label{eq:cumulative-hazard}
\end{equation}
with survival function
\begin{equation}
   \bar{F}(t|\lambda(\cdot)) = 1-F(t|\lambda(\cdot)) = e^{-\Lambda(t)}\ .
   \label{eq:survival-function}
\end{equation}
If observations are censored at time \(\tau\), then the likelihood of \(n\) observations,
\(n_0\) observed and \(n-n_0\) censored, is
\begin{equation}
   L(\lambda(\cdot) | \{t_i\}_{i=1}^n ) = \left[\prod_{i=1}^{n_0} \lambda(t_i) \right]
                                  e^{-\sum_{i=1}^{n_0} \Lambda(t_i) - (n-n_0)\Lambda(\tau)}
   \label{eq:model-likef}
\end{equation}
Our interest is in the estimation of \(\lambda(t)\).

In the construction of Lo and Weng (1989) the hazard rate function is
formed by the following weighted integral
\begin{equation}
  \lambda(t \,|\, G) = \int_{\Theta} \kappa(t \,|\, u) G(du)  
  \label{eq:kernel-mixture}
\end{equation}
where \(G(\cdot)\) is a member of the space of locally finite measures on the
space \(\Theta\) and is the hyperparameter specifying the Gamma Process Prior for \(G\)
(see below for its specification). Three particular forms for the kernel
\(\kappa(t\,|\,u)\) were proposed by Lo and Weng (1989) as follows:
\begin{eqnarray}
  \nonumber
  \kappa(t \,|\, u) &=& I(0 < u < t) \\
  \label{eq:kernelfunctions}
  \kappa(t \,|\, u) &=& I(0 < t < u) \\
  \nonumber
  \kappa(t \,|\, u) &=& I(0 < u < |t-a|), \,\,\, a > 0
\end{eqnarray}
These define an Increasing Failure Rate (IFR), Decreasing Failure
Rate (DFR) and Bathtub failure rate respectively. The latter case we refer to as the
`Lo-Weng Bathtub' (LWB) (see below). We generalise this kernel approach to define three further
bathtub hazard rate functions.

\hypertarget{sec:specificmodels}{%
\subsection{Specific Models}\label{sec:specificmodels}}

We now specify six models using draws from one or more Gamma Process Priors, and
supplemented by additional parameters as necessary.

\textbf{Model 1. Increasing Failure Rate (IFR)} (After Lo and Weng 1989):
\begin{equation}
       \lambda(t|\lambda_0, G) = \lambda_0 + \int_0^t G({\rm d}u)
       \label{eq:ifr}
\end{equation}
Since \(G(\cdot)\) is a non-negative measure, the integral in \eqref{eq:ifr} is a non-decreasing
function of \(t\), giving an increasing hazard rate. A constant
offset background failure rate is given by the parameter \(\lambda_0\geq 0\).

\textbf{Model 2. Decreasing Failure Rate (DFR)} (After Lo and Weng 1989):
\begin{equation}
       \lambda(t|\lambda_0, G) = \lambda_0 + \int_t^\infty G({\rm d}u)
       \label{eq:dfr}
\end{equation}
The integral in \eqref{eq:dfr} is a non-increasing function of \(t\),
giving a decreasing hazard rate, declining to the background rate \(\lambda_0\geq 0\).

\textbf{Model 3. Lo-Weng Bathtub (LWB)} (After Lo and Weng 1989):
\begin{equation}
       \lambda(t|\lambda_0,a,G) = \lambda_0 + \int_0^{|t-a|} G({\rm d}u)
       \label{eq:lwb}
\end{equation}
This model combines the IFR and DFR behaviours of Models 1 and 2.
Here the hazard rate is DFR between \(t=0\) and \(t=a\geq 0\), takes
its minimum value of \(\lambda_0\geq 0\) at \(t=a\) and is IFR thereafter.
The hazard is symmetric around the minimum \(t=a\).

\textbf{Model 4. Superposition Bathtub (SBT)}: To avoid the artificial symmetry of the Lo-Weng Bathtub
Arnold, Chukova, and Hayakawa (2020) proposed
the superposition of two independent DFR and IFR functions.\\
\begin{equation}
       \lambda(t|\lambda_0, G_1, G_2) = \lambda_0 + \int_t^\infty G_1({\rm d}u) + \int_0^t G_2({\rm d}u)
       \label{eq:sbt}
\end{equation}
Here \(G_1\) and \(G_2\) are random measures drawn independently from Gamma Process Priors. The support
for \(G_1\) places mass at early times, and gives DFR behaviour at those times. Late
time IFR behaviour is provided by \(G_2\) which places mass only at later times.

\textbf{Model 5. Mixture Bathtub (MBT)}: Here we adopt the standard specification of a
finite mixture model, and specify the survival function:
\begin{equation}
       \bar{F}(t|\lambda(\cdot|\pi, G_1, G_2)) = \pi \bar{F}_{\rm DFR}(t|G_1) + (1-\pi) \bar{F}_{\rm IFR}(t|G_2)
       \label{eq:mtb}
\end{equation}
Where \(0<\pi<1\) and \(\bar{F}_{\rm DFR}(t|G)\) and \(\bar{F}_{\rm IFR}(t|G)\) are the respective survival functions
from the IFR (Model 1) and DFR (Model 2) cases defined above. Although this specification
does not technically lead to a bathtub hazard rate function (see the discussion in
Section \ref{sec:demonstration} below), it does allow for the high rates of early and late failures.

\textbf{Model 6. Log-convex (LCV)}: If a bathtub hazard rate function is continuous then its derivative is negative
at \(t=0\) and is non-decreasing. This suggests that its derivative could be modelled using the Gamma Process
to give non-negative increments. In order to ensure that the hazard rate function is itself non-negative we apply
these non-negative incremements to the derivative of the \textbf{log} hazard rate:
\begin{equation}
       \frac{d\log\lambda(t|G)}{dt} = w_0 + \int_0^t G({\rm d}u)
       \label{eq:lcvderivative}
\end{equation}
which implies that
\begin{equation}
       \lambda(t|G) = \lambda_0 e^{w_0t + \int_0^t (t-u) G({\rm d}u)}
       \label{eq:lcv}
\end{equation}
Here the constant \(w_0\geq 0\) gives an IFR distribution, but \(w_0<0\) gives a bathtub.

\hypertarget{sec:gammaprocessprior}{%
\subsection{Gamma Process Prior}\label{sec:gammaprocessprior}}

In the hazard rate specifications above we assume \(G\) (or \(G_1\), \(G_2\) in the SBT
and MBT cases) is a draw from a Gamma Process prior \(G\sim\text{GaPP}(\alpha H_0,\beta)\)
with shape parameter \(\alpha>0\), rate parameter \(\beta>0\) and base probability measure
\(H_0(\cdot)\) defined on the space \(\Theta=\mathbb{R}^+\). A draw from the Gamma Process
Prior can be formed by drawing first from a Dirichlet Process Prior
\(\text{DPP}(\alpha H_0)\) and then scaling the resulting weights by an independent
Gamma random variable drawn from \(\gamma\sim\text{Gamma}(\alpha,\beta)\).
To form the draw from the Dirichlet Process Prior we use the stick-breaking
construction of Sethuraman (1994). We note that an alternative stick-breaking
construction for the Gamma Process prior was developed by Roychowdhury and Kulis (2014)
and Roychowdhury and Kulis (2015).

In the gamma-scaled Dirichlet Process Prior method, a draw from the prior
\(\text{GaPP}(\alpha H_0, \beta)\) is
\begin{equation}
  G(du) = \gamma \sum_{k=1}^{\infty} v_k
            \prod_{\ell < k} (1-v_{\ell}) \delta_{\theta_k}(du)
        = \sum_{k=1}^\infty w_k(\gamma,\textbf{v}) \delta_{\theta_k}(du)
  \label{eq:gapp}
\end{equation}
where \(\gamma \sim \mbox{Ga}(\alpha, \beta)\), \(v_k \overset{\text{iid}}{\sim} \mbox{Beta}(1, \alpha)\), and \(\theta_k \overset{\text{iid}}{\sim} H_0\). A
draw from this prior is discrete: there is a countably infinite set of locations
\(\{\theta_k\}_{k=1}^\infty\) and each is associated with a
weight \(\{w_k\}_{k=1}^\infty\). These weights are the product of the Gamma
random draw \(\gamma\) and the usual Dirichlet Process Prior (DPP) weights
\(\{\tilde{w}_k\}_{k=1}^\infty\). These unscaled DPP weights add to 1,
the scaled weights \(w_k\) sum up to the total mass \(\gamma\), and so we have\\
\begin{equation}
  w_k(\gamma,\textbf{v}) 
  = \gamma \tilde{w}_k(\textbf{v}) 
  = \gamma v_k \prod_{\ell < k} (1-v_{\ell})
  \label{eq:weights}
\end{equation}
for \(k=1,2,\ldots\). The weights \(w_k\) are stochastically decreasing in \(k\)
which means that if we truncate the sum in \eqref{eq:gapp} at some sufficiently
large finite \(K\) then the finite sum of the weights \(\sum_{k=1}^K w_k\) will
be very close to \(\gamma\). In practice we draw \(K\) locations \(\theta_k\),
and construct the first \(K-1\) weights using \eqref{eq:weights}. The
weight of the last location is assigned to be \(w_K = \gamma -\sum_{k=1}^{K-1} w_k\)
so that the weights exactly add to \(\gamma\). If \(K\) is sufficiently large then
the effect of this approximation is negligible.

Conditional on this truncation a draw \(G\) from the prior \(G\sim\text{GaPP}_K(\alpha H_0,\beta)\)
can be generated as follows:
\begin{equation}
   \begin{aligned}
   \theta_k | H_0 &\overset{\rm iid}{\sim} H_0 &&\text{for $k=1,2,\ldots,K$}\\
   v_k | \alpha &\overset{\rm iid}{\sim} \text{Be}(1,\alpha) &&\text{for $k=1,2,\ldots,K-1$}\\
   \tilde{w}_k &= v_k \prod_{\ell<k} (1-v_\ell) &&\text{for $k=1,2,\ldots,K-1$}\\
   \tilde{w}_K &= 1-\sum_{k=1}^{K-1} \tilde{w}_k = \prod_{\ell=1}^{K-1} (1-v_\ell)\\
   \gamma &\sim \text{Ga}(\alpha,\beta)\\
   G({\rm d}u) &= \gamma\sum_{k=1}^K \tilde{w}_k \delta_{\theta_k}({\rm d}u)
   \end{aligned}
   \label{eq:gapp-draw}
\end{equation}
Integrals over \(G(\cdot)\) needed in the evaluation of the hazard rate and cumulative hazard rate
are then
\begin{equation}
  \begin{aligned}
  \int_0^t G({\rm d}u) &= \gamma \sum_{k=1}^K \tilde{w}_k I(\theta_k<t)\\
  \int_t^\infty G({\rm d}u) &= \gamma \sum_{k=1}^K \tilde{w}_k I(\theta_k>t)
  \end{aligned}
  \label{eq:gintegral1}
\end{equation}
and
\begin{equation}
  \begin{aligned}
  \int_0^t \int_0^u G({\rm d}v) &= \gamma \sum_{k=1}^K \tilde{w}_k \text{max}(t-\theta_k,0)\\
  \int_0^t \int_u^\infty G({\rm d}v) &= \gamma \sum_{k=1}^K \tilde{w}_k \text{min}(t,\theta_k)
  \end{aligned}
  \label{eq:gintegral2}
\end{equation}
A fully hierarchical Bayesian model for \(G\) can be completed by choosing a form for \(H_0\), with parameters \(\phi\), and specifying suitable priors for \(\alpha\), \(\beta\), \(\gamma\) and \(\phi\). For example if \(H_0=\text{Exp}(\phi)\) then
we can set
\begin{equation}
  \begin{aligned}
   \alpha &\sim& \text{Ga}(a_1,a_2)\\
   \beta  &\sim& \text{Ga}(b_1,b_2)\\
   \phi   &\sim& \text{Ga}(f_1,f_2)
  \end{aligned}
  \label{eq:fullprior}
\end{equation}
for non-negative constants \(\{a_1,a_2,b_1,b_2,f_1,f_2\}\).

\hypertarget{sec:simulation}{%
\section{Hazard Rate Functions and Simulation}\label{sec:simulation}}

If we combine the definitions of the hazard rate functions from \S\ref{sec:specificmodels}
with the implementation of the Gamma Process Priof in \S\ref{sec:gammaprocessprior},
we can obtain expressions for the hazard and cumulative hazard rate functions.
We do this for each model in turn, including suggestions for priors to fully specify the model,
and at the same time provide a protocol for simulation of failure times.\\
We note that if \(U\) is a draw from a Uniform\((0,1)\) distribution,
then the solution \(T\) of the equation
\[
   \Lambda(T) = -\log U
\]
is a draw from the failure distribution with cumulative hazard rate function \(\Lambda(t)\).

In what follows we write Write \(w_k = \gamma\tilde{w}_k\), and also
define \(k^\ast(t) = \text{max}\{k\in (1,\ldots,K)\,:\,\theta^\ast_k\leq t\}\) which implies that
\begin{equation}
    \theta^\ast_{k^\ast(t)} \leq t < \theta^\ast_{k^\ast(t)+1}
\end{equation}
We occasionally need to re-index \(\{(w_k,\theta_k)\}_{k=1}^K\) as
\(\{(w^\ast_k,\theta_k^\ast)\}_{k=1}^K\) such that
\[
    \theta^\ast_0 \equiv 0 < \theta^\ast_1 < \theta^\ast_2 < \ldots < \theta^\ast_K
\]
In doing so we note that if \(H_0\) is chosen to be absolutely continuous then the \(\theta_k\) values
are almost surely distinct.

We also define the following partial sums of the ordered weights and locations:
\begin{eqnarray}
   C^\ast_\ell &=& \sum_{k=1}^\ell w^\ast_k = \gamma - \sum_{k=\ell+1}^K w_k^\ast\\
   D^\ast_\ell &=& \sum_{k=1}^\ell w^\ast_k \theta^\ast_k
\end{eqnarray}

\hypertarget{model-1.-increasing-failure-rate-ifr}{%
\subsubsection*{Model 1. Increasing Failure Rate (IFR)}\label{model-1.-increasing-failure-rate-ifr}}
\addcontentsline{toc}{subsubsection}{Model 1. Increasing Failure Rate (IFR)}

The hazard and cumulative hazard rate functions are:
\begin{eqnarray}
   \lambda(t|\lambda_0,\gamma,\mathbf{\theta},\mathbf{v})
      &=& \lambda_0 + \gamma \sum_{k=1}^K \tilde{w}_k I(\theta_k\leq t)\\
   \Lambda(t|\lambda_0,\gamma,\mathbf{\theta},\mathbf{v})
      &=& \lambda_0 t + \gamma \sum_{k=1}^K \tilde{w}_k \text{max}(0,t-\theta_k)
\end{eqnarray}
If we define \(\Lambda^\ast_k=\Lambda(\theta^\ast_k)\) with \(\theta^\ast_0=0\) and \(\Lambda^\ast_0=0\), then \(\Lambda(t)\) is piecewise linear between the points \(\{(\theta^\ast_k,\Lambda^\ast_k)\}_{k=0}^K\). If \(U\sim\text{Uniform}(0,1)\) then \(T\) is a draw from this distribution if
\begin{eqnarray}
   T = \frac{-\log U + D^\ast_{k^\ast}}{\lambda_0 + C^\ast_{k^\ast}}
\end{eqnarray}
where \(k^\ast = \text{max}\{k\in\{1,\ldots,K\}\,:\,\Lambda^\ast_{k}\leq -\log U\}\). A suitable prior for
\(\lambda_0|G\) is \(\lambda_0|\gamma\sim\text{Exp}(\nu/\gamma)\) for some constant \(\nu>0\).

\hypertarget{model-2.-decreasing-failure-rate-dfr}{%
\subsubsection*{Model 2. Decreasing Failure Rate (DFR)}\label{model-2.-decreasing-failure-rate-dfr}}
\addcontentsline{toc}{subsubsection}{Model 2. Decreasing Failure Rate (DFR)}

The hazard and cumulative hazard rate functions are:
\begin{eqnarray}
   \lambda(t|\lambda_0,\gamma,\mathbf{\theta},\mathbf{v})
      &=& \lambda_0 + \gamma \sum_{k=1}^K \tilde{w}_k I(\theta_k>t)\\
   \Lambda(t|\lambda_0,\gamma,\mathbf{\theta},\mathbf{v})
      &=& \lambda_0 t + \gamma \sum_{k=1}^K \tilde{w}_k \text{min}(t,\theta_k)
\end{eqnarray}
Following the same protocol as in the IFR case:
if we define \(\Lambda^\ast_k=\Lambda(\theta^\ast_k)\) with \(\theta^\ast_0=0\) and \(\Lambda^\ast_0=0\), then \(\Lambda(t)\) is piecewise linear between the points \(\{(\theta^\ast_k,\Lambda^\ast_k)\}_{k=0}^K\). If \(U\sim\text{Uniform}(0,1)\) then \(T\) is a draw from this distribution if
\begin{eqnarray}
   T = \frac{-\log U - D^\ast_{k^\ast}}{\lambda_0 + \gamma - C^\ast_{k^\ast}}
\end{eqnarray}
where \(k^\ast = \text{max}\{k\in\{1,\ldots,K\}\,:\,\Lambda^\ast_{k}\leq -\log U\}\). The same prior for
\(\lambda_0|G\) as in the IFR case can be used: \(\lambda_0|\gamma\sim\text{Exp}(\nu/\gamma)\) for some constant \(\nu>0\).

\hypertarget{model-3.-lo-weng-bathtub-lwb}{%
\subsubsection*{Model 3. Lo-Weng Bathtub (LWB)}\label{model-3.-lo-weng-bathtub-lwb}}
\addcontentsline{toc}{subsubsection}{Model 3. Lo-Weng Bathtub (LWB)}

The hazard and cumulative hazard rate functions are:
\begin{eqnarray}
   \nonumber
   \lambda(t|\lambda_0,a,\gamma,\mathbf{\theta},\mathbf{v})
      &=& \lambda_0 + \gamma \sum_{k=1}^K \tilde{w}_k I(0<\theta_k<|t-a|)\\
      &=& \lambda_0 + 
          \begin{cases}
          \gamma\sum_{k=1}^K \tilde{w}_k I(t<a-\theta_k)\;\; & \text{if $t<a$}\\
          \gamma\sum_{k=1}^K \tilde{w}_k I(t\geq a+\theta_k)\;\; & \text{if $t\geq a$}
          \end{cases}\\
   \Lambda(t|\lambda_0,a,\gamma,\mathbf{\theta},\mathbf{v})
      &=& \lambda_0 t + 
          \begin{cases}
          \gamma\sum_{k=1}^K \tilde{w}_k I(\theta_k<a)\text{min}(t,a-\theta_k)\;\; 
                          & \text{if $t<a$}\\
          \gamma\sum_{k=1}^K \tilde{w}_k 
          \left[\text{max}(0,a-\theta_k)+\text{max}(t-a-\theta_k,0)\right]\;\; & \text{if $t\geq a$}
          \end{cases}\qquad
\end{eqnarray}
We combine the \(\{(\theta_{k},w_k)\}_{k=1}^K\) values with one further location, weight pair: \((a,0)\).\\
We order the locations, and their associated weights, forming the set
\(\{(w^{\ast\ast}_k,\theta^{\ast\ast}_k)\}_{k=1}^{K+1}\), and then compute the cumulative hazard function values
\(\Lambda^{\ast\ast}_k=\Lambda(\theta^{\ast\ast}_k)\). If \(U\sim\text{Uniform}(0,1)\) then \(T\) is a draw
from this distribution if
\begin{eqnarray}
   T = \theta^{\ast\ast}_{k^\ast} 
          + \frac{\theta^{\ast\ast}_{k^\ast+1}-\theta^{\ast\ast}_{k^\ast}}{
                                   \Lambda^{\ast\ast}_{k^\ast+1}-\Lambda^{\ast\ast}_{k^\ast}}
                                   \left(-\log U-\Lambda^{\ast\ast}_{k^\ast}\right)
\end{eqnarray}
where \(k^\ast = \text{max}\{k\in\{1,\ldots,K_1+K_2\}\,:\,\Lambda^{\ast\ast}_{k}\leq -\log U\}\).
The prior \(\lambda_0|\gamma\sim\text{Exp}(\nu/\gamma)\) for some constant \(\nu>0\) is again suitable.

\hypertarget{model-4.-superposition-bathtub-sbt}{%
\subsubsection*{Model 4. Superposition Bathtub (SBT)}\label{model-4.-superposition-bathtub-sbt}}
\addcontentsline{toc}{subsubsection}{Model 4. Superposition Bathtub (SBT)}

We draw \(G_1\) and \(G_2\) from separate Gamma Process priors:
\(G_1\sim\text{GaPP}_K(\alpha_1H_{01},\beta_1)\) and \(G_2\sim\text{GaPP}_K(\alpha_2H_{02},\beta_2)\),
and add a prior for \(\lambda_0|G_2\sim\text{Exp}(\nu/\gamma_2)\).
The hazard and cumulative hazard rate functions are then:
\begin{eqnarray}
   \lambda(t|\lambda_0,\gamma_1,\mathbf{\theta}_1,\mathbf{v}_1,\gamma_2,\mathbf{\theta}_2,\mathbf{v}_2)
      &=& \lambda_0 + \gamma_1 \sum_{k=1}^{K_1} \tilde{w}_{1k} I(\theta_{1k}>t)
                    + \gamma_2 \sum_{k=1}^{K_2} \tilde{w}_{2k} I(\theta_{2k}\leq t)\\
   \Lambda(t|\lambda_0,\gamma_1,\mathbf{\theta}_1,\mathbf{v}_1,\gamma_2,\mathbf{\theta}_2,\mathbf{v}_2)
      &=& \lambda_0 t + \gamma_1 \sum_{k=1}^{K_1} \tilde{w}_{1k} \text{min}(t,\theta_{1k})
                      + \gamma_2 \sum_{k=1}^{K_2} \tilde{w}_{2k} \text{max}(0,t-\theta_{2k}) \qquad
\end{eqnarray}
Extend the IFR/DFR approach: pool the \(\theta_{1k}\) and \(\theta_{2k}\) locations and order them, and their associated weights, forming the set \(\{(w^{\ast\ast}_k,\theta^{\ast\ast}_k)\}_{k=1}^{K_1+K_2}\). Compute the cumulative hazard function values \(\Lambda^{\ast\ast}_k=\Lambda(\theta^{\ast\ast}_k)\). Then if \(U\sim\text{Uniform}(0,1)\) then \(T\) is a draw from this distribution if
\begin{eqnarray}
   T &=& \theta^{\ast\ast}_{k^\ast} 
         + \frac{\theta^{\ast\ast}_{k^\ast+1}-\theta^{\ast\ast}_{k^\ast}}{
                                   \Lambda^{\ast\ast}_{k^\ast+1}-\Lambda^{\ast\ast}_{k^\ast}}
                                   \left(-\log U-\Lambda^{\ast\ast}_{k^\ast}\right)
\end{eqnarray}
where \(k^\ast = \text{max}\{k\in\{1,\ldots,K_1+K_2\}\,:\,\Lambda^{\ast\ast}_{k}\leq -\log U\}\).

\hypertarget{model-5.-mixture-bathtub-mbt}{%
\subsubsection*{Model 5. Mixture Bathtub (MBT)}\label{model-5.-mixture-bathtub-mbt}}
\addcontentsline{toc}{subsubsection}{Model 5. Mixture Bathtub (MBT)}

We draw \(G_1\) and \(G_2\) as in the Superposition Bathtub case, and then the hazard and cumulative hazard rate functions are:
\begin{eqnarray}
   \nonumber
   &&\mbox{}\hspace{-1.8cm}\lambda(t|\pi,
            \lambda_{01},\gamma_1,\mathbf{\theta}_1,\mathbf{v}_1,
            \lambda_{02},\gamma_2,\mathbf{\theta}_2,\mathbf{v}_2)\\
   &=& \frac{f(t|\pi,
                  \lambda_{01},\gamma_1,\mathbf{\theta}_1,\mathbf{v}_1,
                  \lambda_{02},\gamma_2,\mathbf{\theta}_2,\mathbf{v}_2)}{
             \bar{F}(t|\pi,
                  \lambda_{01},\gamma_1,\mathbf{\theta}_1,\mathbf{v}_1,
                  \lambda_{02},\gamma_2,\mathbf{\theta}_2,\mathbf{v}_2)}\\
   \nonumber
   &=& \frac{\pi f_{\rm DFR}(t| \lambda_{01},\gamma_1,\mathbf{\theta}_1,\mathbf{v}_1)
             +(1-\pi) f_{\rm IFR}(t|\lambda_{02},\gamma_2,\mathbf{\theta}_2,\mathbf{v}_2)
   }{\pi \bar{F}_{\rm DFR}(t|\lambda_{01},\gamma_1,\mathbf{\theta}_1,\mathbf{v}_1)
             +(1-\pi) \bar{F}_{\rm IFR}(t|\lambda_{02},\gamma_2,\mathbf{\theta}_2,\mathbf{v}_2)}\\
   \nonumber
   &&\mbox{}\hspace{-1.8cm}\Lambda(t|\pi,\lambda_{01},\gamma_1,\mathbf{\theta}_1,\mathbf{v}_1,
                 \lambda_{02},\gamma_2,\mathbf{\theta}_2,\mathbf{v}_2)\\
   \nonumber
   &=& -\log(\bar{F}(t|\pi,\lambda_{01},\gamma_1,\mathbf{\theta}_1,\mathbf{v}_1,
                           \lambda_{02},\gamma_2,\mathbf{\theta}_2,\mathbf{v}_2))\\
   &=& -\log\left(\pi \bar{F}_{\rm DFR}(t|\lambda_{01},\gamma_1,\mathbf{\theta}_1,\mathbf{v}_1)
            +(1-\pi) \bar{F}_{\rm IFR}(t|\lambda_{02},\gamma_2,\mathbf{\theta}_2,\mathbf{v}_2)\right)\\
   \nonumber
   &=& -\log\left(\pi e^{-\Lambda_{\rm DFR}(t|\lambda_{01},\gamma_1,\mathbf{\theta}_1,\mathbf{v}_1)}
             +(1-\pi) 
             e^{-\Lambda_{\rm IFR}(t|\lambda_{02},\gamma_2,\mathbf{\theta}_2,\mathbf{v}_2)}\right)
\end{eqnarray}
Simulate as a mixture: draw \(c\sim\text{Cat}(\{1,2\};(\pi,1-\pi))\) then if \(c=1\) draw from the DFR component (parameters \((\lambda_{01},\gamma_1,\mathbf{\theta}_1,\mathbf{v}_1)\)), and if \(c=2\) draw from the IFR component (parameters \((\lambda_{02},\gamma_2,\mathbf{\theta}_2,\mathbf{v}_2)\)).

The priors for the two baseline rates \(\lambda_{01}\) and \(\lambda_{02}\) can be
taken to be \(\text{Exp}(\nu/\gamma_1)\) and \(\text{Exp}(\nu/\gamma_2)\) respectively.

\hypertarget{model-6.-log-convex-lcv}{%
\subsubsection*{Model 6. Log convex (LCV)}\label{model-6.-log-convex-lcv}}
\addcontentsline{toc}{subsubsection}{Model 6. Log convex (LCV)}

The hazard and cumulative hazard rate functions implied by \eqref{eq:lcv} are:
\begin{eqnarray}
   \lambda(t|\lambda_0,w_0,\gamma,\mathbf{\theta},\mathbf{v}) &=& \lambda_0 \exp\left(w_0 t + \sum_{k=1}^K w_k  
       \text{max}(0,t-\theta_k)\right)\\
       \nonumber
   \Lambda(t|\lambda_0,w_0,\gamma,\mathbf{\theta},\mathbf{v}) &=& \sum_{\ell=0}^{k^\ast(t)-1} \lambda_0 e^{-D^\ast_\ell}
                      \left\{\frac{e^{C^\ast_{0\ell}\theta^\ast_{\ell+1}}
                            -e^{C^\ast_{0\ell}\theta^\ast_{\ell}}}{C^\ast_{0\ell}}\right\}\\
                  & & + \lambda_0 e^{- D^\ast_{k^\ast(t)}}
                      \left\{\frac{e^{C^\ast_{0k^\ast(t)}t}
                          -e^{C^\ast_{0k^\ast(t)}\theta^\ast_{k^\ast(t)}}}{C^\ast_{0k^\ast(t)}}\right\}
\end{eqnarray}
where we are using the re-indexed set of weights and locations \(\{(w^\ast_k,\theta^\ast_k)\}_{k=1}^K\), and
where
\begin{eqnarray}
   C^\ast_{0\ell} = w_0 + C^\ast_\ell = \sum_{k=0}^\ell w^\ast_k
\end{eqnarray}
Note that \(C^\ast_{0k}\) is non-decreasing in \(k\), and can in some circumstances take the value zero. Where
\(C^\ast_{0k}\) is zero the two quantities in the braces in the expression for \(\Lambda(t|\lambda_0,w_0,\gamma,\theta,{\bf v})\) should be replaced by
\[
    \theta^\ast_{k+1}-\theta^\ast_{k}
\]
and
\[
    t - \theta^\ast_{k^\ast(t)}
\]
respectively.

Again following the same protocol as in the IFR case
we define \(\Lambda^\ast_k=\Lambda(\theta^\ast_k)\) with \(\theta^\ast_0=0\) and \(\Lambda^\ast_0=0\). Then in the interval
\(\theta^\ast_{k^\ast(t)}\leq t < \theta^\ast_{k^\ast(t)}\) the function \(\Lambda\) has the exponential form
\begin{eqnarray}
   \Lambda(t|\lambda_0,w_0,\gamma,\mathbf{\theta},\mathbf{v})
   &=&
   \Lambda^\ast_{k^\ast(t)} + \frac{\lambda_0 e^{- D^\ast_{k^\ast(t)}}}{C^\ast_{0l^\ast}}
                      \left\{e^{C^\ast_{0k^\ast(t)}t}
                           -e^{C^\ast_{0k^\ast(t)}\theta^\ast_{k^\ast(t)}}\right\}
\end{eqnarray}
It follows that
If \(U\sim\text{Uniform}(0,1)\) then \(T\) is a draw from this distribution if
\begin{eqnarray}
   T = \frac{1}{C^\ast_{0k^\ast}}\log\left[
         e^{C^\ast_{0k^\ast}\theta^\ast_{k^\ast}}
       + C^\ast_{0k^\ast}\frac{-\log U - \Lambda^\ast_{k^\ast}}{\lambda_0 e^{-D^\ast_{k^\ast}}}
       \right]
\end{eqnarray}
where \(k^\ast = \text{max}\{k\in\{1,\ldots,K\}\,:\,\Lambda^\ast_{k}\leq -\log U\}\). In the case
where \(C^\ast_{0k^\ast}=0\) instead we have:
\[
   T = \theta^\ast_{k^\ast} + \frac{-\log U - \Lambda^\ast_{k^\ast}}{\lambda_0 e^{-D^\ast_{k^\ast}}}
\]
Priors are needed for \(\lambda_0\) and \(w_0\). Suitable priors are
\begin{eqnarray}
   \log\lambda_0 &\sim& \text{Normal}(0,(\gamma/\nu)^2)\\
   w_0 &\sim& \text{Normal}(0,(\gamma/\nu)^2) 
\end{eqnarray}

\hypertarget{sec:demonstration}{%
\section{Demonstration}\label{sec:demonstration}}

\begin{figure}

{\centering \includegraphics{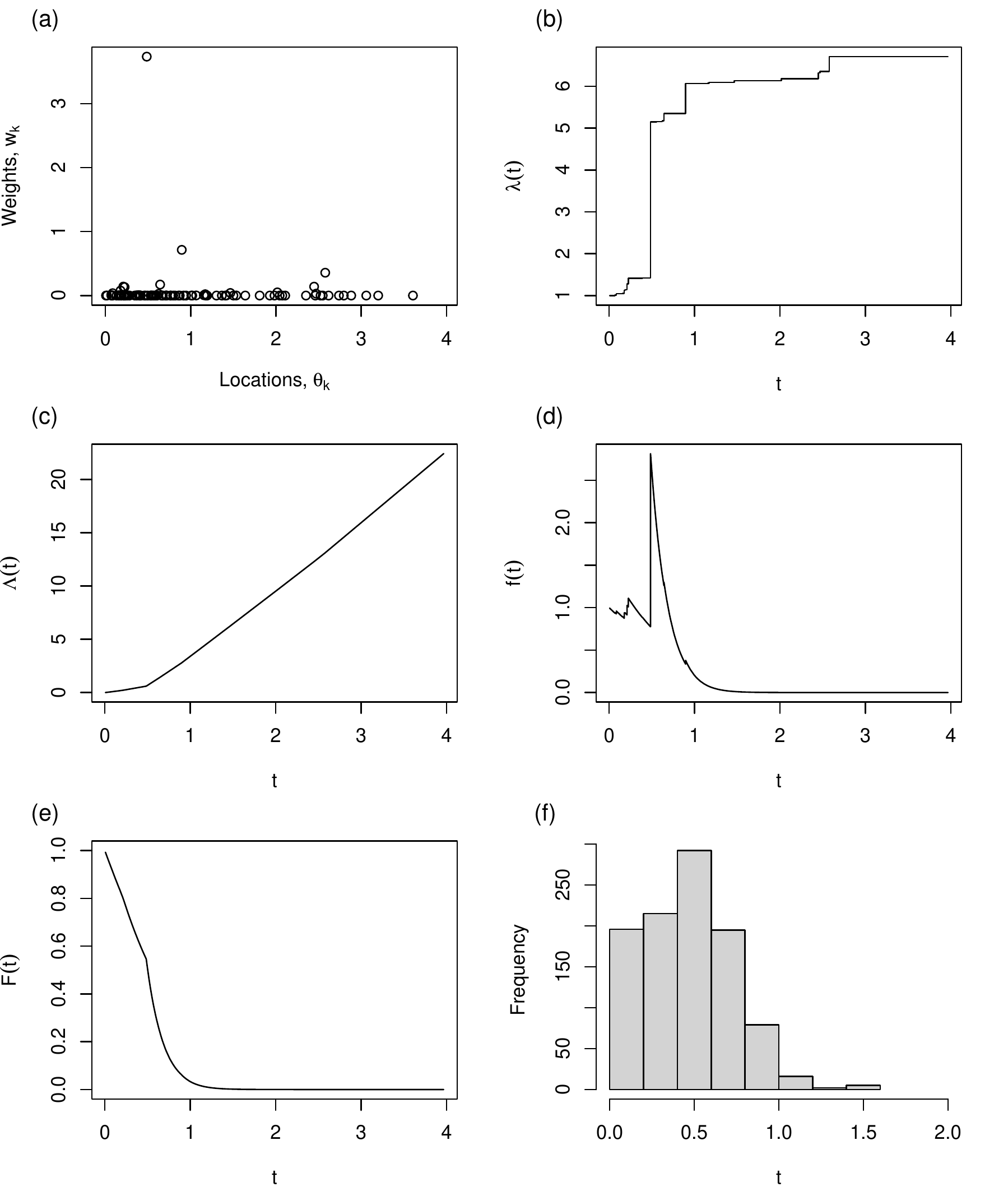} 

}

\caption{{\bf Model 1. Increasing Failure Rate (IFR):} A single draw from the prior, and a sample of simulated data. (a) Locations and weights $\{(\theta_k,w_k)\}$; (b) Hazard rate function $\lambda(t)$; (c) Cumulative hazard rate function $\Lambda(t)$; (d) Probability density $f(t)$; (e) Survival Function $\bar{F}(t)$; (f) Sample of $n=1000$ failure time observations.}\label{fig:ifr-demo}
\end{figure}

\begin{figure}

{\centering \includegraphics{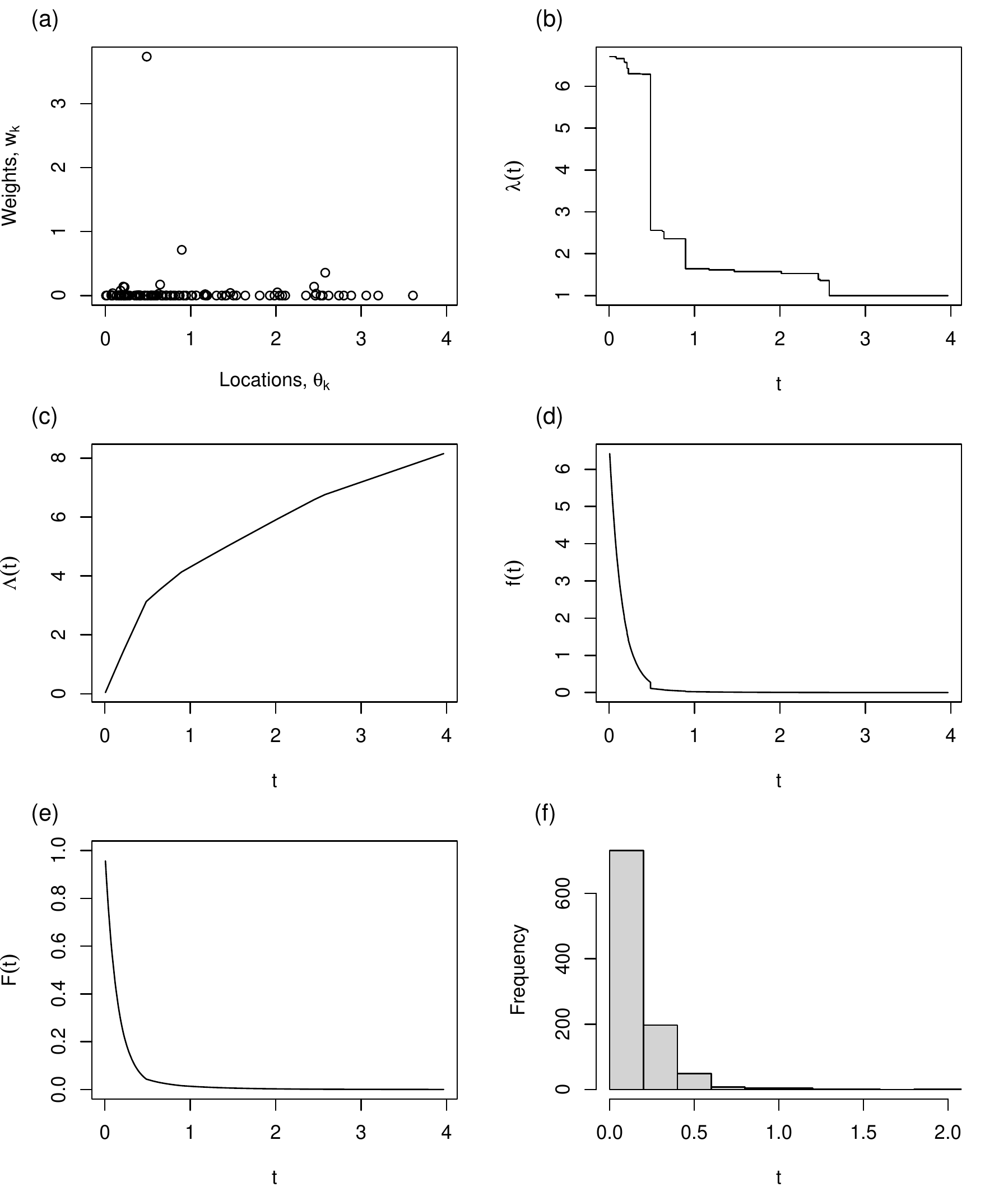} 

}

\caption{{\bf Model 2. Decreasing Failure Rate (DFR):} A single draw from the prior, and a sample of simulated data. (a) Locations and weights $\{(\theta_k,w_k)\}$; (b) Hazard rate function $\lambda(t)$; (c) Cumulative hazard rate function $\Lambda(t)$; (d) Probability density $f(t)$; (e) Survival Function $\bar{F}(t)$; (f) Sample of $n=1000$ failure time observations.}\label{fig:dfr-demo}
\end{figure}

\begin{figure}

{\centering \includegraphics{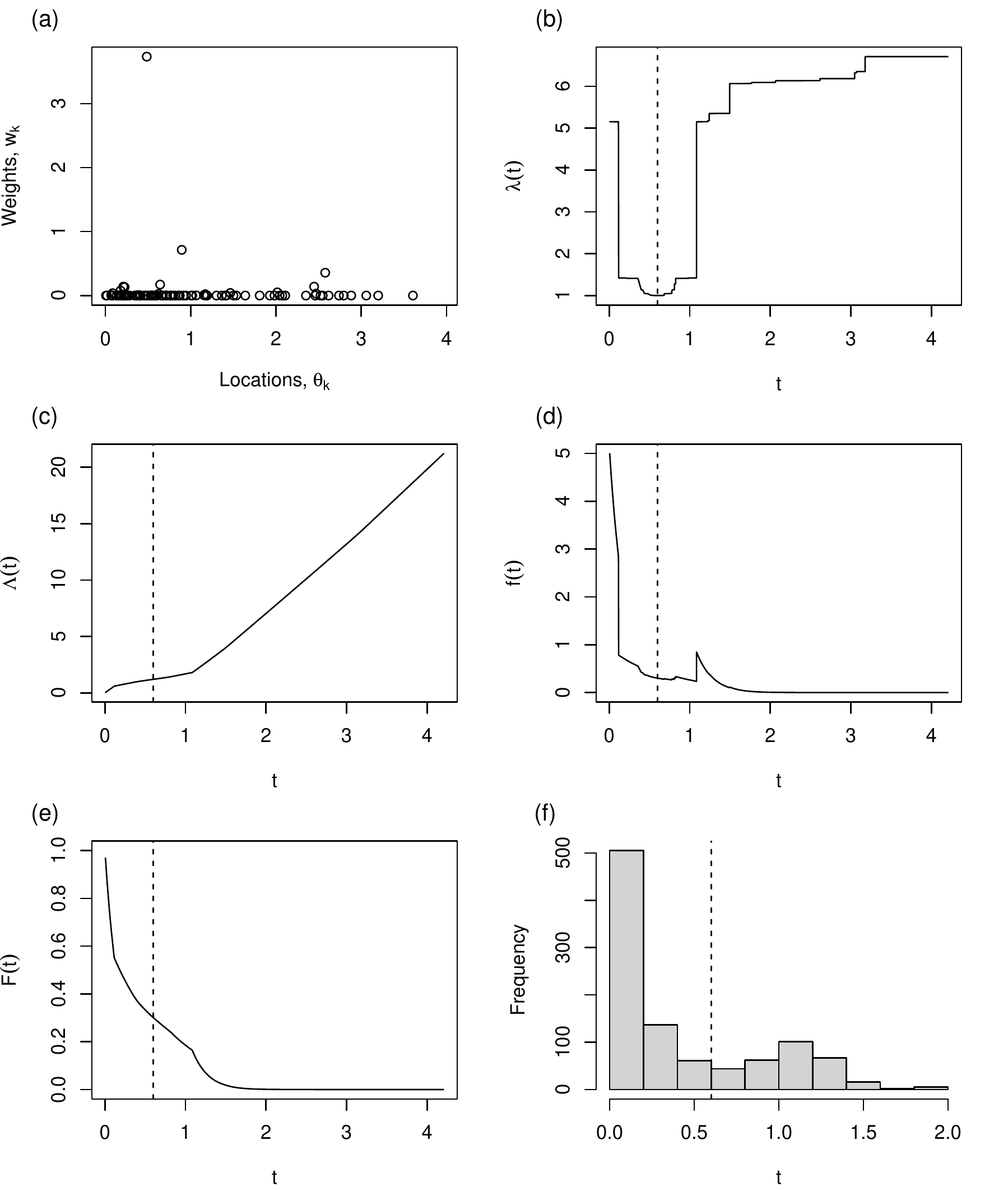} 

}

\caption{{\bf Model 3. Lo-Weng Bathtub (LWB):} A single draw from the prior, and a sample of simulated data. (a) Locations and weights $\{(\theta_k,w_k)\}$; (b) Hazard rate function $\lambda(t)$; (c) Cumulative hazard rate function $\Lambda(t)$; (d) Probability density $f(t)$; (e) Survival Function $\bar{F}(t)$; (f) Sample of $n=1000$ failure time observations. The location of the inflection point $a$ is shown by vertical dashed lines.}\label{fig:lwb-demo}
\end{figure}

\begin{figure}

{\centering \includegraphics{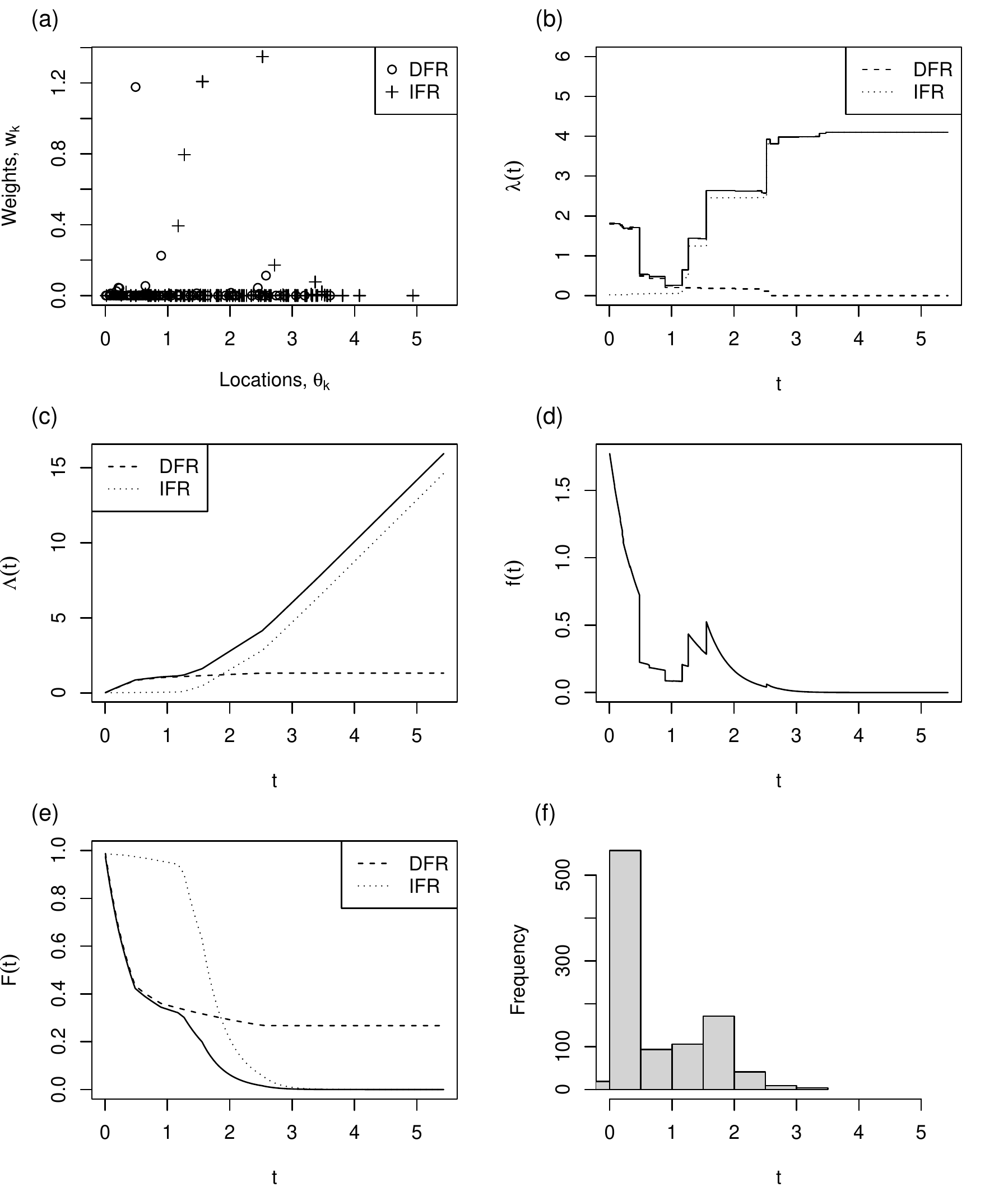} 

}

\caption{{\bf Model 4. Superposition Bathtub (SBT):} A single draw from the prior, and a sample of simulated data. (a) Locations and weights $\{(\theta_k,w_k)\}$ with the DFR component weights shown as open circles, and the IFR component as '+' symbols; (b) Hazard rate function $\lambda(t)$; (c) Cumulative hazard rate function $\Lambda(t)$; (d) Probability density $f(t)$; (e) Survival Function $\bar{F}(t)$; (f) Sample of $n=1000$ failure time observations.}\label{fig:sbt-demo}
\end{figure}

\begin{figure}

{\centering \includegraphics{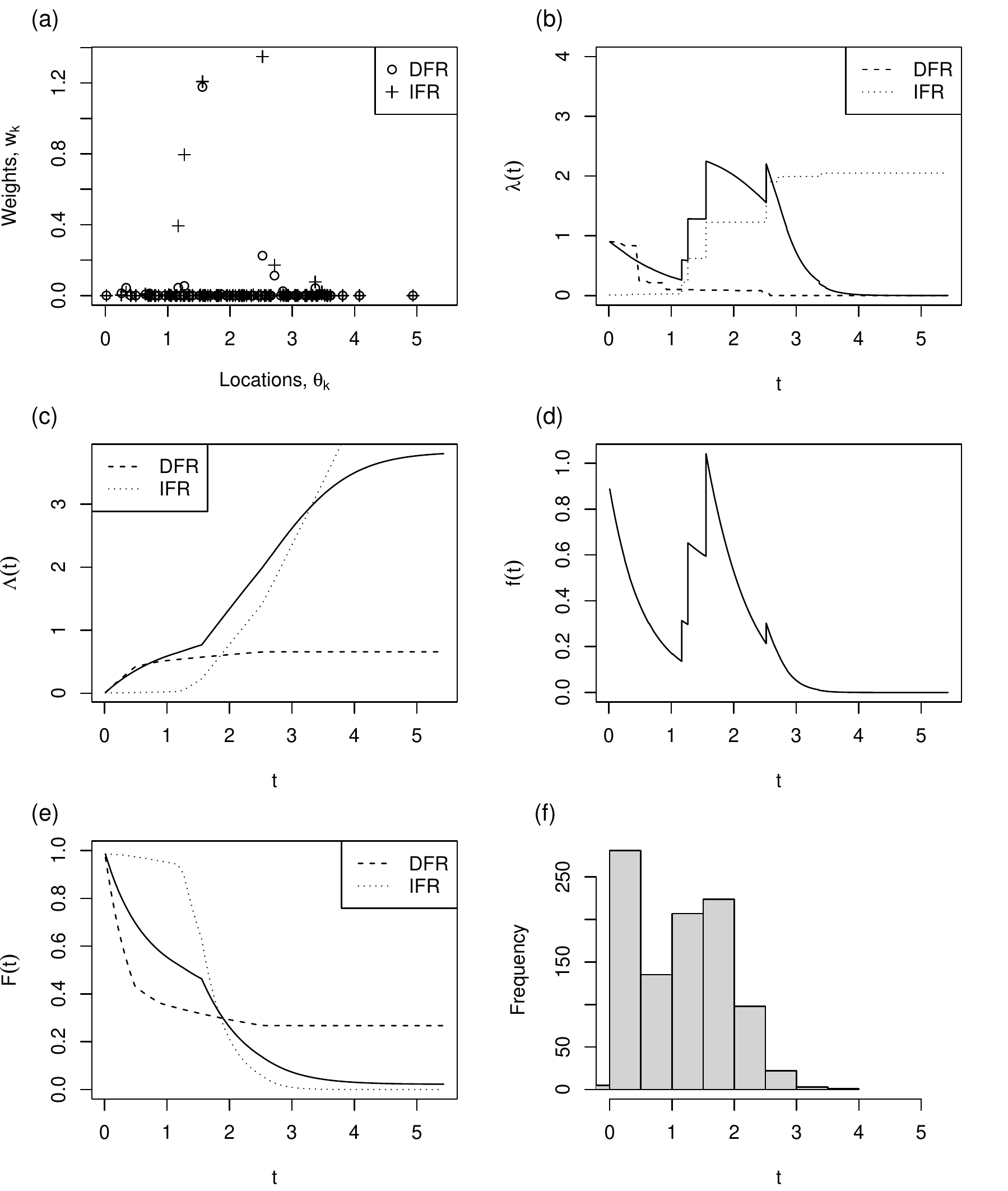} 

}

\caption{{\bf Model 5. Mixture Bathtub (MBT):} A single draw from the prior, and a sample of simulated data. (a) Locations and weights $\{(\theta_k,w_k)\}$ with the DFR component weights shown as open circles, and the IFR component as '+' symbols; (b) Hazard rate function $\lambda(t)$; (c) Cumulative hazard rate function $\Lambda(t)$; (d) Probability density $f(t)$; (e) Survival Function $\bar{F}(t)$; (f) Sample of $n=1000$ failure time observations.}\label{fig:mbt-demo}
\end{figure}

\begin{figure}

{\centering \includegraphics{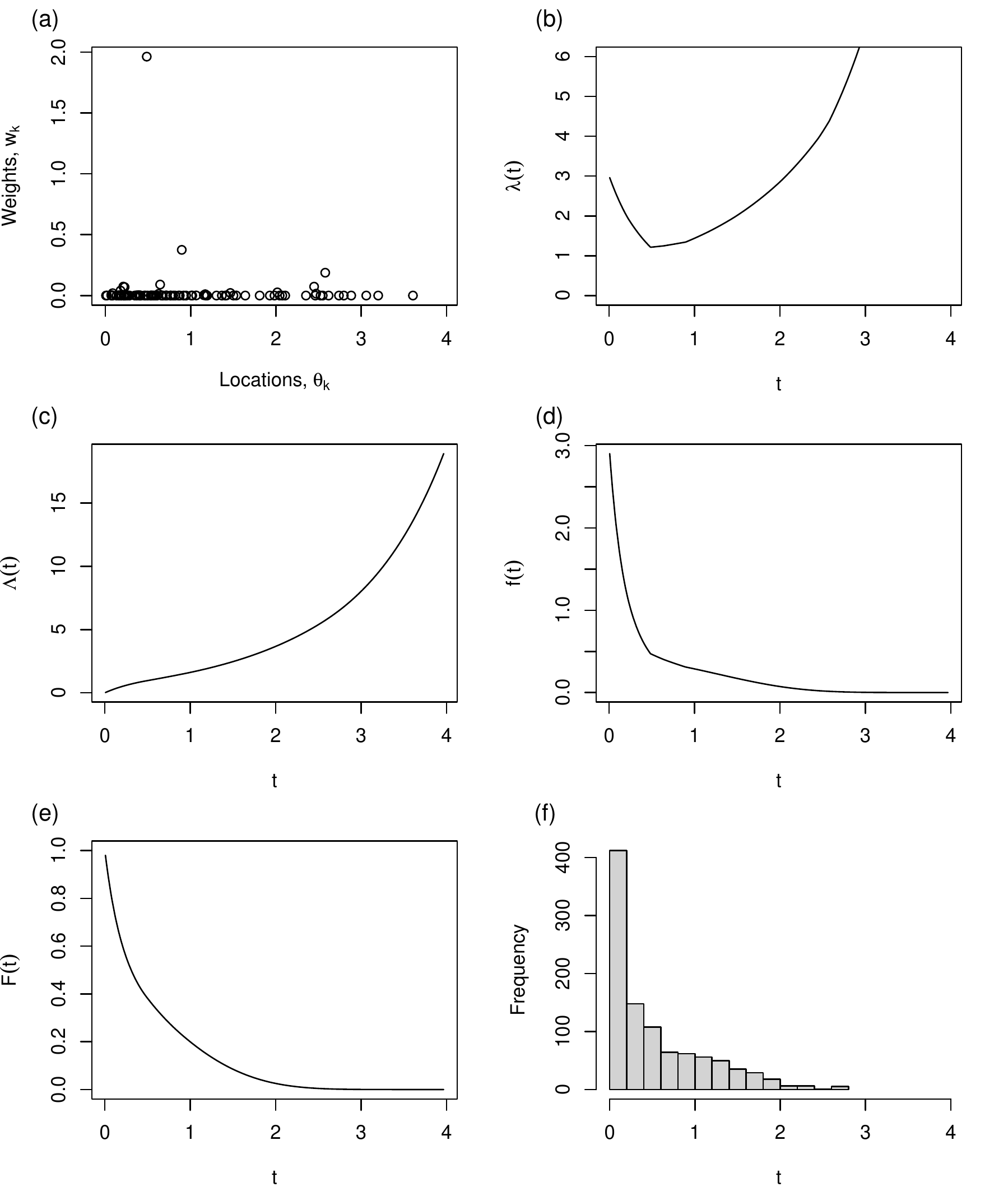} 

}

\caption{{\bf Model 6. Log Convex (LCV):} A single draw from the prior, and a sample of simulated data. (a) Locations and weights $\{(\theta_k,w_k)\}$; (b) Hazard rate function $\lambda(t)$; (c) Cumulative hazard rate function $\Lambda(t)$; (d) Probability density $f(t)$; (e) Survival Function $\bar{F}(t)$; (f) Sample of $n=1000$ failure time observations.}\label{fig:lcv-demo}
\end{figure}

We now demonstrate the properties of the six models specified in Section \ref{sec:model}. Figures
\ref{fig:ifr-demo}-\ref{fig:lcv-demo} show example draws from each of the models, and in each case
we show the location/weight pairs \(\{(\theta_k,w_k)\}\), the hazard rate function
\(\lambda(t)\), the cumulative hazard rate \(\Lambda(t)\), the density function \(f(t)=\lambda(t)\exp(-\Lambda(t))\)
and the survival function \(\bar{F}(t)=\exp(-\Lambda(t))\). We also show a histogram of \(n=1000\) random
draws from the failure time distribution.

\begin{figure}

{\centering \includegraphics{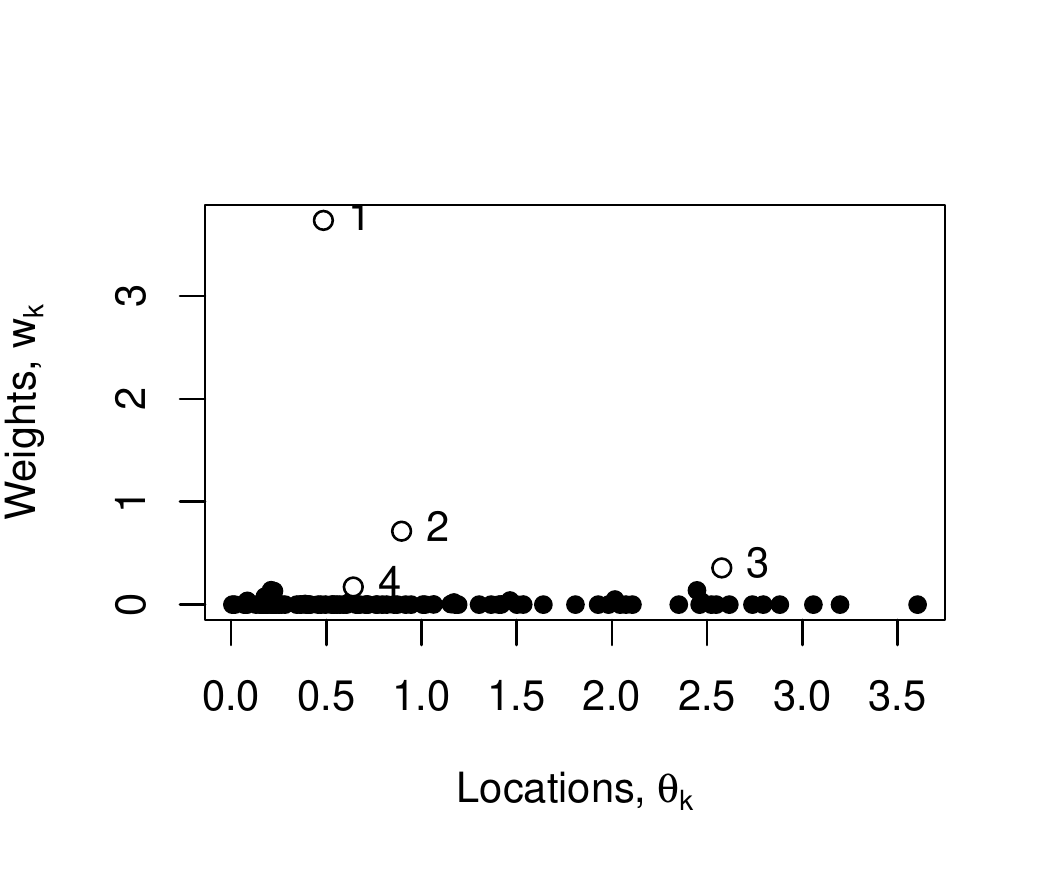} 

}

\caption{A draw from the Gamma Process Prior.  The locations and weights are plotted as filled circles, except the first four points $k=1,2,3,4$ are labelled and shown as white circles.}\label{fig:points-weights}
\end{figure}

We note that any draw from the Gamma Process prior is concentrated on a set of discrete points, and that
the weights \(w_k\) decrease stochastically in \(k\). This is what enables the truncation of the otherwise
infinite sums in \S\ref{sec:gammaprocessprior}. Consider a draw \(G\) from
\(G\sim\text{GaPP}_K(\alpha H_0,\beta)\) as given in \eqref{eq:gapp-draw}. We set \(\alpha=3\),
\(\beta=1\) and choose the baseline distribution \(H_0\) as \(\text{Exp}(\phi)\) with \(\phi=1\). Figure
\ref{fig:points-weights} shows the first \(K=100\) location/weight pairs \(\{(\theta_k,w_k)\}_{k=1}^K\). The
first 4 weights (labelled individually in Figure \ref{fig:points-weights}) account for
87.2\% of the total weight, and the first 40
account for all but 0.0001\% of the total weight.

The discrete nature of the draw \(G\) is visible in Figures \ref{fig:ifr-demo}-\ref{fig:lcv-demo} as multiple
step changes in the hazard rate function \(\lambda(t)\), and in sharp peaks and troughs in the failure time
distribution \(f(t|\lambda(\cdot))\).

To aid comparison we have used the same single draw \(G\) from Figure \ref{fig:points-weights} for the IFR, DFR,
LWB and LCV models, and have used this same draw (though rescaled) as \(G_1\) in the SBT and MBT models.
The draw \(G_2\) in the SBT and MBT models has the same \(\alpha=3\) and \(\beta=1\) parameter
values, but the baseline failure rate distribution \(H_0\) is a Normal\((2,1^2)\).

The Increasing Failure Rate case (Model 1, IFR, Figure \ref{fig:ifr-demo}) is a simple case where the
hazard rate \(\lambda(t)\) increases by the corresponding weight value at each value of \(\theta\). The hazard
rate thus jumps upwards at these discrete points and is constant between them.
The point with the largest weight in Figure \ref{fig:ifr-demo}(a) is responsible
for the single largest jump in \(\lambda(t)\), and corresponds to the sharp peak in the failure time
density \(f(t)\) in Figure \ref{fig:ifr-demo}(d), and also to the sudden sharp decline in the survival function
\(\bar{F}(t)\) in Figure \ref{fig:ifr-demo}(e).

The Decreasing Failure Rate case (Model 2, DFR, Figure \ref{fig:dfr-demo}) shows a hazard rate
function \(\lambda(t)\) the opposite behaviour to the IFR case: namely at each weight location the hazard rate
jumps downwards. With the hazard rate so large at early times the failure time distribution is
more strongly concentrated towards zero than in the IFR case.

For the Lo-Weng Bathtub (Model 3, LWB, Figure \ref{fig:lwb-demo})
the hazard rate function has reflectional symmetry about \(t=a\), which we have set to be at \(a=0.6\).
There are downward (DFR) jumps
of \(w_k\) at times \(a-\theta_k\) and equal upward (IFR) jumps at times \(a+\theta_k\). This bathtub hazard rate
is responsible for the bimodal failure time distribution \(f(t)\) in Figure \ref{fig:lwb-demo}(d).

The Superposition Bathtub (Model 4, SBT, Figure \ref{fig:sbt-demo}) has a hazard rate function that
is the linear combination of a DFR and IFR hazard rate function. In Figure \ref{fig:sbt-demo}(b) the
sum of these two monotonic hazard rate functions leads to the bathtub shape of \(\lambda(t)\), and again
to the bimodal failure time distribution \(f(t)\). Figure \ref{fig:sbt-demo}(a) shows the two separate
sets of weights and locations for the two draws \(G_1\sim\text{GaPP}_K(\alpha_1H_{01},\beta_1)\) and
\(G_2\sim\text{GaPP}_K(\alpha_2H_{02},\beta_2)\) where \(H_{01}\) is an exponential distribution and
\(H_{02}\) is a Normal distribution.

As we noted earlier, the Mixture Bathtub (Model 5, MBT, Figure \ref{fig:mbt-demo}) case does
not have an overall bathtub hazard function, but does show the characteristic U-shape at early
times (a decrease followed by an increase). This leads to the bimodal failure time distribution
\(f(t)\). In this example the two components have been given equal weight in the mixture: \(\pi=0.5\).

The Log Convex case (Model 6, LCV, Figure \ref{fig:lcv-demo}) differs somewhat from the others due to
the random measure \(G\) contributing linearly to the logarithm of the hazard rate function. The
corresponding hazard rate function \(\lambda(t)\) is piecewise curved rather than being piecewise constant.
The example shown in Figure \ref{fig:lcv-demo} shows a failure time distribution with a monotonically
decreasing rather than bimodal density. With appropriate choice of parameters this
monotonic behaviour can be
seen in all of Models 3-6, and is a consequence of a large integrated hazard at early times making it
less likely that failures occur at late times. The increasing hazard in such situations serves to
shorten the tail of \(f(t)\), rather than creating a mode at late times.

\hypertarget{sec:conclusions}{%
\section{Conclusions}\label{sec:conclusions}}

This paper lays out the details of simulation for four different bathtub
hazard rate functions (Models 3-6), based on draws from the Gamma Process Prior. These
non-parametric specifications are highly flexible, and include only
minimal assumptions about the form of the hazard rate function.

The examples in Section \ref{sec:demonstration} demonstrate the range of behaviours that
the various models can exhibit, and we expect to be able to find suitable models
to match many real data sets.

The next step is thus inference from data, which we intend to address in a future
paper. Arnold, Chukova, and Hayakawa (2020) have
demonstrated how inference can be carried out for the IFR model using MCMC samplers
in a fully Bayesian framework. Some of the parameter updates are Gibbs however
updates of the weights \(\{w_k\}\) and locations \(\{\theta_k\}\) require Metropolis-Hastings
proposals. Arnold, Chukova, and Hayakawa (2020)
noted that their samplers may need further development to ensure
that they converge efficiently, and made various suggestions for improvements
which may be especially relevant in the case of censored data.

We anticipate that the model selection question, choosing which among these models
is best, may be carried out using a criterion such as WAIC (Watanabe 2010), with model
checking using posterior predictive distributions.

\hypertarget{acknowledgements}{%
\subsection*{Acknowledgements}\label{acknowledgements}}
\addcontentsline{toc}{subsection}{Acknowledgements}

This work was supported by: Waseda University,
Grant for Special Research Projects (2018K-383); JSPS KAKENHI Grant-in-Aid for
Scientific Research (C) Grant Number 18K04621; Waseda Institute for Advanced
Study Visiting Scholars 2018; FY2018 and FY2022 Grant Program for Promotion of
International Joint Research, Waseda University. Fulbright New Zealand:
Fulbright Scholar Award 2018.

\hypertarget{references}{%
\section*{References}\label{references}}
\addcontentsline{toc}{section}{References}

\hypertarget{refs}{}
\begin{CSLReferences}{1}{0}
\leavevmode\vadjust pre{\hypertarget{ref-AlAbbasi-etal:2019}{}}%
Abbasi, Jamal N. Al, Mundher A. Khaleel, Moudher Kh. Abdal‑hammed, Yue Fang Loh, and Gamze Ozel. 2019. {``{A new uniform distribution with bathtub‑shaped failure rate with simulation and application}.''} \emph{Mathematical Science} 13: 105--14.

\leavevmode\vadjust pre{\hypertarget{ref-Arnold_Chukova_Hayakawa_APARM:2020}{}}%
Arnold, Richard, Stefanka Chukova, and Yu Hayakawa. 2020. {``{Nonparametric Bayesian Analysis of Hazard Rate Functions using the Gamma Process Prior}.''} In \emph{{2020 Asia-Pacific International Symposium on Advanced Reliability and Maintenance Modeling (APARM)}}.

\leavevmode\vadjust pre{\hypertarget{ref-DykstraLaud:1981}{}}%
Dykstra, R. L., and Purushottam Laud. 1981. {``{A Bayesian Non-parametric Approach to Reliability}.''} \emph{Annals of Statistics} 9: 356--67.

\leavevmode\vadjust pre{\hypertarget{ref-Ferguson:1974}{}}%
Ferguson, Thomas S. 1974. {``{Prior Distributions on Spaces of Probability Measures}.''} \emph{The Annals of Statistics} 2: 615--29.

\leavevmode\vadjust pre{\hypertarget{ref-Glaser:1980}{}}%
Glaser, Ronald E. 1980. {``{Bathtub and Related Failure Rate Characterizations}.''} \emph{Ournal of the American Statistical Association} 75: 667--72.

\leavevmode\vadjust pre{\hypertarget{ref-Hayakawa.etal:2001}{}}%
Hayakawa, Yu, Jonathan Zukerman, Sue Paul, and Tony Vignaux. 2001. {``{Bayesian non-parametric testing of constant versus non-decreasing hazard rates}.''} In \emph{System and Bayesian Reliability}, edited by Yu Haykawa, Telba Z. Irony, and Min Xie, 5:391--406. Series on Quality, Reliability and Engineering Statistics. Singapore: World Scientific.

\leavevmode\vadjust pre{\hypertarget{ref-HoLo:2001}{}}%
Ho, Man-wai, and A. Y. Lo. 2001. {``{Bayesian non-parametric estimation of a monotone hazard rate}.''} In \emph{System and Bayesian Reliability}, edited by Yu Haykawa, Telba Z. Irony, and Min Xie, 5:301--14. Series on Quality, Reliability and Engineering Statistics. Singapore: World Scientific.

\leavevmode\vadjust pre{\hypertarget{ref-Mutairi-etal:2021}{}}%
Iqbal, Alya Al Al MutairiMuhammad Zafar, Muhammad Zafar Iqbal, Zeeshan Arshad, Badr Alnssyan, Hazem Al-Mofleh, and Ahmed Z. Afify. 2021. {``{A New Extended Model with Bathtub Shaped Failure Rate: Properties, Inference, Simulation, and Applications}.''} \emph{Mathematics} 9: 2024:1--32. \url{https://doi.org/10.3390/math9172024}.

\leavevmode\vadjust pre{\hypertarget{ref-KaplanMeier:1958}{}}%
Kaplan, E. L., and Paul Meier. 1958. {``{Nonparametric Estimation from Incomplete Observations}.''} \emph{Journal of the American Statistical Association} 53: 457--81.

\leavevmode\vadjust pre{\hypertarget{ref-Kingman:1967}{}}%
Kingman, J. F. C. 1967. {``{Completely random measures}.''} \emph{Pacific Journal of Mathematics} 21: 59--78.

\leavevmode\vadjust pre{\hypertarget{ref-KumarKlefsjoGranholm:1989}{}}%
Kumar, U., B. Klefsjö, and S. Granholm. 1989. {``{Reliability investigation for a fleet of load haul dump machines in a Swedish mine}.''} \emph{Reliability Engineering and System Safety} 26: 341--61.

\leavevmode\vadjust pre{\hypertarget{ref-LoBrunnerChan:1998}{}}%
Lo, A. Y., L. J. Brunner, and A. T. Chan. 1998. {``{Weighted Chinese restaurant processes and Bayesian mixture models (Revision 1.1)}.''} Department of Information; Systems Management, Hong Kong University of Science; Technology, Hong Kong: Hong Kong University of Science; Technology.

\leavevmode\vadjust pre{\hypertarget{ref-LoWeng:1989}{}}%
Lo, A. Y., and C-S. Weng. 1989. {``{On a class of Bayesian non-parametric estimates II. Hazard rate estimates}.''} \emph{Annals of the Institute of Statistical Mathematics} 41: 221--45.

\leavevmode\vadjust pre{\hypertarget{ref-MullerQuintanaJaraHanson:2015}{}}%
Müller, Peter, Fernando Andrés Quintana, Alejandro Jara, and Tim Hanson. 2015. \emph{{Bayesian Nonparametric Data Analysis}}. New York, NY: springer.

\leavevmode\vadjust pre{\hypertarget{ref-Paisley:2010}{}}%
Paisley, J. 2010. {``{A Simple Proof of the Stick-Breaking Construction of the Dirichlet Process}.''} Department of Computer Science, Princeton University, Princeton, NJ: Princeton University. \url{http://www.columbia.edu/~jwp2128/Teaching/E6892/papers/SimpleProof.pdf}.

\leavevmode\vadjust pre{\hypertarget{ref-Paisley.Blei.Jordan:2012}{}}%
Paisley, J., D. M. Blei, and M. I. Jordan. 2012. {``{Stick-Breaking Beta Processes and the Poisson Process}.''} In \emph{Proceedings of the 15th International Conference on Artificial Intelligence and Statistics (AISTATS) 2012, La Palma, Canary Islands}, 850--58.

\leavevmode\vadjust pre{\hypertarget{ref-Paisley.Zaas.Woods.Ginsburg.Carin:2010}{}}%
Paisley, J., A. Zaas, C. W. Woods, G. S. Ginsburg, and L. Carin. 2010. {``{A Stick-Breaking Construction of the Beta Process}.''} In \emph{Proceedings of the 26th International Conference on Machine Learning, Haifa, Israel, 2010}, 1--8.

\leavevmode\vadjust pre{\hypertarget{ref-PengLiuWang:2016}{}}%
Peng, Chong, Guangpeng Liu, and Lun Wang. 2016. {``{Piecewise modelling and parameter estimation of repairable system failure rate}.''} \emph{SpringerPlus} 5: 1477:1--14.

\leavevmode\vadjust pre{\hypertarget{ref-Phadia:2016}{}}%
Phadia, Eswar G. 2016. \emph{{Prior Processes and Their Applications: Nonparametric Bayesian Estimation}}. 2nd ed. New York, NY: Springer.

\leavevmode\vadjust pre{\hypertarget{ref-Roychowdhury.Kulis:2014}{}}%
Roychowdhury, Anirban, and Brian Kulis. 2014. {``{Gamma Processes, Stick-Breaking, and Variational Inference}.''} \url{https://arxiv.org/abs/1410.1068}.

\leavevmode\vadjust pre{\hypertarget{ref-Roychowdhury.Kulis:2015}{}}%
---------. 2015. {``{Gamma Processes, Stick-Breaking, and Variational Inference}.''} In \emph{Proceedings of the 18th International Conference on Artificial Intelligence and Statistics (AISTATS)}, 800--808.

\leavevmode\vadjust pre{\hypertarget{ref-Sethuruman:1994}{}}%
Sethuraman, Jayaram. 1994. {``{A constructive definition of Dirichlet priors}.''} \emph{Statistica Sinica} 4: 639--50.

\leavevmode\vadjust pre{\hypertarget{ref-Watanabe:2010}{}}%
Watanabe, Sumio. 2010. {``{Asymptotic Equivalence of Bayes Cross Validation and Widely Applicable Information Criterion in Singular Learning Theory}.''} \emph{Journal of Machine Learning Research} 11: 3571--94.

\end{CSLReferences}

\end{document}